# A Dynamic Programming Framework for Combinatorial Optimization Problems on Graphs with Bounded Pathwidth

Mugurel Ionut Andreica[1]
[1]Polytechnic University of Bucharest, mugurel.andreica@cs.pub.ro

*Abstract*-In this paper we present an algorithmic framework for solving a class of combinatorial optimization problems on graphs with bounded pathwidth. The problems are NP-hard in general, but solvable in linear time on this type of graphs. The problems are relevant for assessing network reliability and improving the network's performance and fault tolerance. The main technique considered in this paper is dynamic programming.

## I. INTRODUCTION

Network reliability analysis and the improvement of the network's performance and fault tolerance are issues of great interest to the networking community. Careful network analysis and testing, based on relevant reliability metrics, can point out network vulnerabilities which could severely impact network performance, while improving the network's fault tolerance can help eliminate some of these problems. The network can be modeled as an undirected graph, with network nodes as vertices and network links as edges. The vertices and edges may have several parameters associated to them, like cost, radius (in the case of wireless networks), latency, bandwidth and many others. We believe that some of the properties of the corresponding graph model can be used in order to define effective network reliability metrics and for improving the network's performance level and degree of fault tolerance. In this paper, we present efficient algorithms for computing some important properties and solving combinatorial optimization problems for the class of graphs with bounded pathwidth. We focus here only on the algorithms, whose efficiency is important especially in the case of large graphs (like those encountered in practical situations), and leave other aspects for future work. All these algorithms are presented as part of a generic framework, which can be further extended with algorithms not considered in this paper. The rest of the paper is structured as follows. In Section II we present formal definitions of the concepts used in the rest of the paper. In Section III we present a generic dynamic programming framework for solving combinatorial optimization problems on graphs with bounded pathwidth. In Section IV we present several clear examples regarding the usage of the framework and in Section V we present related work. Finally, in Section VI we conclude and mention directions for future research.

## II. GRAPHS WITH BOUNDED PATHWIDTH

The pathwidth of an undirected graph is a number which reflects the resemblance of the graph's structure to a path – the lower the pathwidth, the closer the graph "looks" like a path. A path decomposition of a graph G is a path D, with nodes $D_1$, $D_2$, ..., $D_P$ (in the order they lie on the path), having the following properties:

- each node $D_i$ corresponds to a subset of $nv(i) \geq 0$ vertices of G (we will denote the subset by $D_i$, too)
- any two adjacent vertices of the graph G, u and v, belong together to at least one subset $D_i$
- each vertex u of G belongs to at least one subset $D_i$ and if u belongs to two subsets $D_i$ and $D_k$, then it also belongs to all the subsets in between $D_i$ and $D_k$ (the subsets which contain a vertex u form a sub-path of D)

The width of the path decomposition is defined as $pw_D = \max\{nv(1), \ldots, nv(P)\} - 1$. The minimum value of $pw_D$ of a path decomposition of the graph is called the graph's pathwidth. Finding a path decomposition with minimum width is an NP-hard problem, but in many practical situations, a decomposition whose width is bounded by a constant can be easily found. Moreover, some efficient algorithms for finding path decompositions of small width have been developed [4].

The pathwidth concept is strongly related to the notion of treewidth, which was introduced by Robertson and Seymour [1]. The treewidth captures the degree of similarity of a graph's structure to a tree. Many NP-hard problems can be solved in polynomial time on graphs whose pathwidth (or treewidth) are bounded by a constant. These algorithms are usually based on the dynamic programming technique and have a time complexity of the form $O(f(pw) \cdot n)$, where $f(pw)$ is a function which is exponential in the width of the path decomposition pw, and n is the number of vertices of the graph. The algorithms make use of a path decomposition of the graph. In order to simplify the algorithms, we will introduce the concept of nice path decompositions. The nodes (subsets) of a nice path decomposition are of the following two types:

- *Introduce node*: If $D_i$ is an introduce node, then $D_i = D_{i-1} \cup \{x\}$, where x is a vertex which does not belong to $D_{i-1}$ (the introduced vertex). $D_1$ is an introduce node consisting of just one vertex.
- *Forget node*: If $D_i$ is a forget node, then $D_i = D_{i-1} \setminus \{x\}$, where x is a vertex which belongs to $D_{i-1}$, but not to $D_i$ (the forgotten vertex). $D_P$ is a forget node with $nv(P)=0$.

Any path decomposition can be easily transformed into a nice path decomposition with O(n) nodes in O(n) time [5]. All the algorithms in the subsequent section will consider that a nice path decomposition is already known.

## III. A Dynamic Programming Framework

Dynamic programming algorithms traverse the nodes of the given nice path decomposition in order and for each node i they compute a table $T_i$. The size of the table $T_i$ is exponential in the number of vertices of the subset $D_i$. Each entry of the table contains a state S and a value v, i.e. $T_i[S]=v$. S is the state of the vertices in $D_i$ and is usually composed of one or several values for each vertex in $D_i$. v is the value of the optimization function, restricted to the vertices in

$$UD_i = \bigcup_{j=1}^{i} D_j$$

and considering that the vertices in $D_i$ are in state S. $T_i[S]$ is computed based on the values $T_{i-1}[S']$, for some states S' which are *compatible* with the state S. The definition of state compatibility depends on the actual problem solved (just like the definition of the state itself). Each state obeys several structural rules, which depend on the problem. We will call the states which violate some of these rules *intermediate* states. These states will need to be *normalized* into valid states. The set of all valid states of a node $D_i$ is called $VS_i$.

Within the proposed generic algorithm, we will iterate through all the states for node $D_{i-1}$ and expand these states into valid states for the node $D_i$. The expansion function will depend on the actions that we can perform (which are problem dependent) and on node $D_i$'s type (Introduce or Forget node). In the end, the solution will be found in one of the entries of the table $T_P$, considering only states belonging to a subset of *valid final states*. We are only interested in finding the value of an optimization function, not the states of the graph vertices leading to the optimal value. However, these states can easily be computed from the tables stored for each node of the path decomposition (by going back from node P to node 1). The generic dynamic programming algorithm is given below:

**Generic Dynamic Programming Algorithm:**
**compute** $T_1[S]$, **for all** *states S* **in** $VS_1$
**for** *i=1* **to** *P-1* **do**
  **for all** *states S* **in** $VS_{i+1}$ **do**
    $T_{i+1}[S]=uninitialized$
  **for** *S* **in** $VS_i$, *such that* $T_i[S] \neq uninitialized$ **do**
    **for** *action* **in** *setOfActions(i+1)* **do**
      // S' is an intermediate state (not necessarily valid)
      // C is the (new) value of the optimization function
      *(S',C,ok)=***expandState***(S, i, action)*
      **if** *(ok=true)* **then**
        *S''=***normalize***(S')*
        **if** *((***better***(C, T_{i+1}[S''])) or (T_{i+1}[S'']=uninitialized))* **then**
          $T_{i+1}[S'']=C$
*OPT=uninitialized*
**for** *S* **in** *setOfValidFinalStates()* **do**
  **if** (**better**($T_P[S]$, *OPT*)) or (*OPT=uninitialized*) **then**
    *OPT=$T_P[S]$*
**return** *OPT*

From an implementation point of view, the states for each node will be generated in an array of states, which can be traversed easily. When reading or writing a value $T_i[S]$, we need to know the index of state S in the array of states (between 1 and the total number of states). The most efficient way to do this is to use two hash functions ($hash_1$ and $hash_2$). $hash_1$ will generate a unique hash value for each state S (no collisions are allowed). This value will be stored in a hash table, together with the state index. The hash table will use the $hash_2$ function and permits some collisions. The pseudocode below illustrates the use of this approach.

**generateStates(i):** // generates all the states for node $D_i$
*stateIndex=0*
**for each** *state S generated* **do**
  $h_1=$**hash$_1$***(S)*
  *stateIndex=stateIndex+1*
  *hashTable[i].put(S, stateIndex)* // hashTable[i] uses hash$_2$()
**getStateIndex(S, i)**
  **return** *hashTable[i].get(***hash$_1$***(S))*

Since we are discussing efficiency, we should note that the sets of states of two nodes $D_i$ and $D_j$ will differ only if $nv(i) \neq nv(j)$. This suggests that we could generate the states only for each distinct value of the number of vertices (there are only $pw_D+1$ such values) and not for each node.

## IV. Combinatorial Optimization Problems

In this section we will present several combinatorial optimization problems which can be solved using the generic framework presented in the previous section. The problems have practical applications to network reliability analysis and fault tolerance and performance improvement.

### A. Coloring a graph with a fixed number of colors

We are given a graph G together with a nice path decomposition of the graph. We have to assign to each vertex of the graph a color from the set {1,2,…,C}, such that any two vertices connected by an edge are assigned different colors.

This is one of the simplest problems, in which the function which needs to be computed is a binary function. We need to decide if a coloring exists or not. If it exists, the vertex colors can be derived from the tables stored at each node of the path decomposition. Furthermore, we can use the solution to this problem in a binary (or linear) search algorithm, in order to find the minimum number of colors required to color the graph.

The state of the vertices of a node $D_i$ of the path decomposition has the form $S=(c_1, c_2, …, c_{nv(i)})$, where $c_i \in \{1,…,C\}$ is the color of the $i^{th}$ vertex in the subset $D_i$. We will, occasionally, denote by S[i] the $i^{th}$ component of the state S. We will consider the vertices of a node $D_i$ ordered as $v_{i,1}, v_{i,2}, …, v_{i,nv(i)}$. If $D_i$ is an Introduce node, then we will consider that the introduced vertex is $v_{i,nv(i)}$. We will maintain these vertex ordering assumptions in all the other problems considered in this section. An entry $T_i[S]$ has one of the values *true* or *uninitialized*, meaning that there exists (does not exist) a coloring of the vertices in $UD_i$, such that the vertices in $D_i$ are colored according to the state S. Node $D_1$ contains only a single vertex, so we will assign $T_1[S]=true$, for all the states S in $VS_1$. The set of actions which can be performed for expanding a state S of the node i into a state S' of the node i+1 depends on the type of the node $D_{i+1}$. If $D_{i+1}$ is an Introduce node, the set of actions consists of coloring the introduced node in every possible color; if it is a Forget node, only a "forget" action exists. We will now define all the functions required to turn the generic algorithm from Section II into a solution to the problem.

**setOfActions(i):**
  **if** ($D_i$ *is an Introduce node*) **then**
    **return** *{ (Col, 1), (Col, 2), ..., (Col, C) }*
  **else return** *{ Forget }*

**updateCost(S, i, C):** // auxiliary function, used by *expandState*
  let $S=(c_1, c_2, ..., c_{nv(i)})$
  **for** *j=1* **to** *nv(i)-1* **do**
    **if** (($v_{i,j}$ *and* $v_{i,nv(i)}$ *are adjacent*) **and** ($c_j=c_{nv(i)}$)) then
      **return** *( (), 0, false)*
  **return** (*S, 1, true*)

**expandState(S, i, action):**
  **if** ($D_i$ *is an Introduce node*) **then**
    *(Col, cx)=action* // cx is the color assigned to the new vertex
    $S'=(c_1, c_2, ..., c_{nv(i)-1}, c_{nv(i)}=cx)$, **where** $S=(c_1, c_2, ..., c_{nv(i)-1})$
    **return** updateCost(*S', i, $T_{i-1}[S]$*)
  **else**
    $v_{i-1,j}$ = the "forgotten" node
    $S'=(c_1, c_2, ..., c_{j-1}, c_{j+1}, ..., c_{nv(i-1)})$, **where** $S=(c_1, c_2,..., c_{nv(i-1)})$
    **return** (*S',1,true*)

**normalize(S):**
  **return** *S*

**better(cost$_1$, cost$_2$):**
  **if** ($cost_1=1$) **then return** *true*
  **else return** *false*

**setOfValidFinalStates():**
  **return** $VS_P$ // all the states of $D_P$ are valid final states

It is obvious that the *expandState* function is the most important one in the algorithm and this will be the case with each problem we will consider. In this function, the selected action is performed and the validity of the resulting intermediate state is checked. The complexity of the algorithm is $O((pw+1) \cdot C^{pw+1} \cdot P)$, considering that the path decomposition has width pw. Since P is O(n) and $(pw+1) \cdot C^{pw+1}$ is bounded by a constant, the time complexity of the algorithm is linear.

*B. Coloring a graph with a fixed number of colors – improved state definition*

The improvement of the previous solution consists in reducing the number of states. It is obvious that, given a valid coloring of the graph's vertices, we can relabel the colors differently and still get a valid coloring. For instance, if C=3 and we have two vertices a and b colored with colors 3 and 2, respectively, we can relabel the colors such that vertex a is colored with 1 and vertex b is colored with 2. This suggests that the colors of a state S should form a partition and obey the following rules:

- $c_1=1$
- $c_i \in \{1,..., m_{i-1}+1\}$, where $m_{i-1} = \max_{1 \leq j \leq i-1} \{c_j\}$

With these rules, the state S' returned by the *expandState* function may not be a valid state. Therefore, we will have to define the normalize function differently:

**normalize(S):**
  $S'=S$, where $S=(c_1, ..., c_K)$
  *counter=0*
  *newlabel={0,0,...,0}* // K zeroes
  **for** *i=1* **to** *K* **do**
    **if** (*newlabel[S[i]]=0*) **then**
      *counter=counter+1*
      *newlabel[S[i]]=counter*
    *S'[i]=newlabel[S[i]]*
  **return** *S'*

The *normalize* function relabels the colors of a state S such that they obey the structural rule. The number of states is greatly reduced. For instance, for C=7 and a node $D_i$ with nv(i)=9, the number of states is equal to the number of partitions of a set with 9 elements into at most 7 parts, which is 21,110. Before, the number of states was $9^7$= 4,782,969.

*C. Coloring a graph with a fixed number of colors in order to minimize penalties due to coloring conflicts*

This problem is similar to the previous one, except that a valid coloring is not necessarily required. Each graph edge (u,v) has an associated penalty value *pen(u,v)*. If the vertices u and v are assigned the same color, then the penalty pen(u,v) will be paid. The optimization function consists of minimizing the sum of paid penalties. For this problem, we will keep the same state definition as in the previous case, the same sets of actions and the same valid final states. We will have to slightly modify the *expandState* function, by redefining the auxiliary function *updateCost*, and the *better* function. $T_i[S]$ now represents the minimum penalty paid such that all vertices in $UD_i$ are colored and the vertices in $D_i$ are colored according to the state S. $T_1$ will be initialized with 0 for every possible state.

**updateCost(S, i, C):**
  *C'=C*
  **for** *j=1* **to** *nv(i)-1* **do**
    **if** ((*adjacent*($v_{i,j}, v_{i,nv(i)}$)) **and** (*S[j]=S[nv(i)]*)) **then**
      *C'=C'+pen*($v_{i,j}, v_{i,nv(i)}$)
  **return** (*S,C',true*)

**better(cost$_1$, cost$_2$):**
  **if** ($cost_1<cost_2$) **then return** *true*
  **else return** *false*

No other changes are necessary in order to solve this problem, which has applications to frequency assignment in wireless networks. If we want to solve a slightly different version of the problem, in which we try to minimize the maximum penalty paid instead of the sum of penalties, we only have to change the additive operator in the *updateCost* function with the *max* operator (*C'=max{C', pen($v_{i,j}, v_{i,nv(i)}$)}*). A different solution to this modified problem consists of binary searching the cost to be paid. When the cost C is fixed, we can ignore all the edges with a penalty lower than (or equal to) C and we would now have to solve a normal coloring problem.

*D. Minimum Path Cover*

A path cover of a graph G consists of a union of disjoint paths Path$_1$, Path$_2$, …, Path$_{PC}$, which cover all the vertices of G. More formally:

- Path$_i$=$p_{i,1}, p_{i,2}, …, p_{i,npv(i)}$, where npv(i) is the number of vertices on path i and two consecutive vertices $p_{i,j}$ and $p_{i,j+1}$ are connected by an edge
- Each vertex of the graph G belongs to exactly one path

We are interested in minimizing the number of paths in the path cover (PC). Note that this problem contains finding a Hamiltonian path as a particular case and is NP-hard in general. The state for a node $D_i$ with $nv(i)$ vertices is defined as $S=(s_1, s_2, \ldots, s_{nv(i)})$. $s_j$ is the state of the $j^{th}$ vertex of node $D_i$ (considering the same ordering as before). $s_j$ can take one of the following values:

- $s_j=-1$ implies that vertex $v_{i,j}$ has degree zero in the path cover (it does not have any neighbors)
- $s_j=0$ implies that vertex $v_{i,j}$ has degree two in the path cover (it has two neighbors => it lies inside a path)
- $s_j>0$ implies that $v_{i,j}$ has degree 1 in the path cover and is one of two endpoints of a path; $s_j$ is the path's identifier

If $s_j>0$, there can be at most one other vertex $v_{i,k}$ with $s_k=s_j$ (the other endpoint of the same path). It is also possible that the other endpoint does not belong to $D_i$ (it was "left behind"). In this problem, $T_i[S]$ represents the minimum number of paths which cover all the vertices in $UD_i$, considering that the vertices in $D_i$ are in state S. The only valid state for $D_1$ is $S=(-1)$, with $T_1[S]=1$. We will define next all the functions required by the generic dynamic programming algorithm.

**setOfActions(i):**
  **if** ($D_i$ is an Introduce node) **then**
    $actionSet=\{ newPath \}$
    **for** $j=1$ to $nv(i)-1$ **do**
      **if** ($adjacent(v_{i,j}, v_{i,nv(i)})$) **then**
        $actionSet = actionSet \cup \{(extendPath, v_{i,j})\}$
    **for** $j=1$ to $nv(i)-2$ **do**
      **for** $k=j+1$ to $nv(i)-1$ **do**
        **if** ($adjacent(v_{i,j}, v_{i,nv(i)})$ **and** $adjacent(v_{i,k}, v_{i,nv(i)})$) **then**
          $actionSet = actionSet \cup \{(connectPaths, v_{i,j}, v_{i,k})\}$
    **return** $actionSet$
  **else return** $\{ Forget \}$

**expandState(S, i, action):**
  **let** $S=(s_1, s_2, \ldots, s_{nv(i)-1})$
  **if** ($D_i$ is an Introduce node) **then**
    **if** ($action=newPath$) **then**
      $S'=(s_1, \ldots, s_{nv(i)-1}, -1)$
      **return** $(S', T_{i-1}[S]+1, true)$
    **else if** ($action=(extendPath, v_{i,j})$) **then**
      **if** ($s_j=0$) **then return** $( (), +Infinity, false)$
      **else if** ($s_j>0$) **then**
        $S'=(s_1, \ldots, s_{j-1}, 0, s_{j+1}, \ldots, s_{nv(i)-1}, s_j)$
        **return** $(S', T_{i-1}[S], true)$
      **else** // $s_j=-1$
        $pid=max\{0, s_1, s_2, \ldots, s_{nv(i)-1}\}+1$
        $S'=(s_1, \ldots, s_{j-1}, pid, s_{j+1}, \ldots, s_{nv(i)-1}, pid)$
        **return** $(S', T_{i-1}[S], true)$
    **else**
      **let** $action=(connectPaths, v_{i,j}, v_{i,k})$
      **if** (($s_j=-1$) **and** ($s_k=-1$)) **then**
        $pid=max\{0, s_1, s_2, \ldots, s_{nv(i)-1}\}+1$
        $S'=(s_1, \ldots, s_{j-1}, pid, s_{j+1}, \ldots, s_{k-1}, pid, s_{k+1}, \ldots, s_{nv(i)-1}, 0)$
        **return** $(S', T_{i-1}[S]-1, true)$
      **else if** (($s_j=-1$) **and** ($s_k>0$)) **then**
        $S'=(s_1, \ldots, s_{j-1}, s_k, s_{j+1}, \ldots, s_{k-1}, 0, s_{k+1}, \ldots, s_{nv(i)-1}, 0)$
        **return** $(S', T_{i-1}[S]-1, true)$
        // the case ($s_j>0$) and ($s_k=-1$) is treated in a similar manner
      **else if** (($s_j>0$) **and** ($s_k>0$) **and** ($s_j \ne s_k$)) **then**
        $S'=(s_1, \ldots, s_{nv(i)-1}, 0)$
        // relabel the other endpoint of one of the two paths
        **find** $k'$ such that $s_{k'}=s_k$
        **if** ($k'$ exists) **then** $S'[k']=s_j$
        $S'[j]=S'[k]=0$
        **return** $(S', T_{i-1}[S]-1, true)$
      **else** // we tried to connect the endpoints of the same path
        **return** $( (), +Infinity, false)$
  **else**
    $v_{i-1,j}=$ the "forgotten" node
    $S'=(s_1, s_2, \ldots, s_{j-1}, s_{j+1}, \ldots, s_{nv(i-1)})$
    **return** $(S', T_{i-1}[S], true)$

The *normalize* function is almost the same as for the previous problem, except that all the values $s_j$ equal to -1 or 0 are left unchanged. Only the path ids are relabeled, such that they form a partition into classes in which every class contains at most two vertices. The *better* and *setOfValidStates* functions are also maintained. The time complexity of the algorithm is $O(pw^3 \cdot max\{|VS_i|\} \cdot n)$, where $|VS_i|$ is exponential in pw.

### E. Minimum Cycle Cover

A minimum cycle cover has a definition which is almost identical to the minimum path cover, except that it consists of cycles, i.e. paths in which the first and last vertex must also be connected by an edge (this excludes paths composed of only one or two vertices). There are some minor adjustments which need to be made. First, $T_i[S]$ will represent the minimum number of *closed* cycles which cover all the vertices in $UD_i$ and the state of the vertices in $D_i$ is S. Furthermore, all the vertices in $UD_i \backslash D_i$ belong to an open or closed cycle. Both endpoints of the *open* cycles (i.e. paths) must belong to $D_i$ (except when the open cycle contains only one vertex). This changes the state definition slightly, in the sense that if $S=(s_1, \ldots, s_{nv(i)})$ is the state of the vertices of the node $D_i$ and $s_j>0$ for a vertex $v_{i,j}$, there must exist a vertex $v_{i,k}$ with $s_k=s_j$. The set of valid final states contains only one state $S_{final}=(0, 0, \ldots, 0)$, i.e. when there are no open cycles left. The set of actions for an *Introduce node* is extended in order to contain actions of type *(closeCycle, $v_{i,j}, v_{i,k}$)*, where $v_{i,j}$ and $v_{i,k}$ must be the two endpoints of an open cycle. The cost value returned by the expandState function is always $T_{i-1}[S]$, except when an open cycle is closed, in which case the cost will be $T_{i-1}[S]+1$. If $D_i$ is a *Forget node* in the *expandState* function, the algorithm must first verify that the state of the "forgotten" node $v_{i-1,j}$ in S is $S[j]=0$ (i.e. it is not the endpoint of an open cycle). If $S[j] \ne 0$ the returned tuple will be $((), +Infinity, false)$. For lack of space, the described changes will not be presented in pseudocode. We draw attention to the fact the Minimum Cycle Cover contains the Hamiltonian Cycle as a particular case and it is NP-hard in general.

### F. k-Replica Placement

We are given an undirected graph with n vertices together with a nice path decomposition with small pathwidth pw. We want to select k distinct vertices of the graph and place a replica of some popular content in them. The cost of selecting a vertex i is *csel(i)*. If two vertices u and v which are adjacent to one another are selected, then we will also need to pay a penalty cost *pen(u,v)*. We are interested in paying the minimum total cost for placing the k replicas. The state definition we will use

is the following: for a node $D_i$, a state S has the form $(s_1, ..., s_{nv(i)}, x)$, where:
- $s_j=1$ if $v_{i,j}$ was selected for placing a replica
- $s_j=0$ if $v_{i,j}$ was not selected for placing a replica
- x is the total number of vertices selected (so far)

The set of actions of an Introduce node consists of two actions *{ Select, Do Not Select }* and that of a Forget node will be the same as before (*{Forget}*). We will show the main functions required by the framework. The *normalize* function will not be presented (because all the intermediate states will be valid) and the valid final states will be only those with x=k.

**updateCost(S, i, C):**
  C'=C
  **for** *j=1* **to** *nv(i)-1* **do**
   **if** ((adjacent($v_{i,j}$, $v_{i,nv(i)}$) **and** (S[j]=S[nv(i)]=1)) **then**
    C'=C'+pen($v_{i,j}$, $v_{i,nv(i)}$)
  **return** (S,C',true)

**expandState(S, i, action):**
  let $S=(s_1, s_2, ..., s_{nv(i)-1}, x)$
  **if** ($D_i$ is an Introduce node) **then**
   **if** (action=Select) **then**
    **if** (x=k) **then**
     **return** ((), +Infinity, false)
    $S'=(s_1, s_2, ..., s_{nv(i)-1}, 1, x+1)$
    **return** updateCost(S', i, $T_{i-1}$[S]+csel($v_{i,nv(i)}$))
   **else** // action = Do Not Select
    $S'=(s_1, s_2, ..., s_{nv(i)-1}, 0, x)$
    **return** (S', $T_{i-1}$[S], true)
  **else**
   $v_{i-1,j}$ = the "forgotten" node
   $S'=(s_1, s_2, ..., s_{j-1}, s_{j+1}, ..., s_{nv(i-1)}, x)$
   **return** (S',$T_{i-1}$[S],true)

We will use the same *better* function as in the minimum penalty coloring. The time complexity of the algorithm is $O(k \cdot 2^{pw} \cdot n)$. We can introduce several variations to this problem, like defining penalty or profit values for each pair of adjacent vertices (u,v), where u is a selected vertex and v is not. These changes would require a different *updateCost* function.

*G. Maximum Leaf Weighted Spanning Tree*

We are given an undirected graph G with n>1 vertices and a nice path decomposition of G. Each vertex i has an associated weight w(i). We want to find a spanning tree of G such that the total weight of the leaves (vertices of degree 1) of the spanning tree is maximum. This is a more general version of the well-known Maximum Leaf Spanning Tree problem, which is NP-hard in general. The states for a node $D_i$ have the following form $((cid_1, deg_1), (cid_2, deg_2), ..., (cid_{nv(i)}, deg_{nv(i)}))$. $cid_j$ is the identifier of the connected component to which vertex $v_{i,j}$ belongs. $deg_j$ is the degree of vertex $v_{i,j}$ in its connected component. We are only interested in the values 0, 1 and 2 (if $v_{i,j}$ has degree greater than 2, we will keep its value at 2). All the connected components are trees. The identifiers of the connected components form a partition, so they must obey the same rules as in the coloring problems presented previously. Every connected component must have at least one representative vertex in the set of vertices of the currently processed node i (i.e no connected component is "left behind"). When introducing a node, the actions are of three types: *newComponent*, *addAsLeaf*  and *connectComponents*. When forgetting a vertex $v_{i-1,j}$, we must check that at least one other vertex $v_{i-1,k}$ with the same cid still exists; otherwise, the connected component of $v_{i-1,j}$ would be "left behind". The only valid state is the one in which all the vertices are in the same connected component (all the cids are 1). $T_i$[S] will represent the maximum total weight of the leaves of the connected components, such that very vertex in $UD_i$ belongs to a component and the vertices in $D_i$ are in state S. We will only present the *setOfActions* and *expandState* functions, because the others can be derived from the problems presented previously in this section.

**setOfActions(i):**
 **if** ($D_i$ is an Introduce node) **then**
  actionSet={ newComponent }
  **for** *j=1* **to** *nv(i)-1* **do**
   **if** (adjacent($v_{i,j}$,$v_{i,nv(i)}$)) **then**
    actionSet = actionSet ∪ {(addAsLeaf, $v_{i,j}$)}
  **for all sets** $S=\{v_{i,j1}, v_{i,j2}, ..., v_{i,jk}\}$ with k>1 **do**
   **if** ($v_{i,nv(i)}$ is adjacent to all the vertices in S) **then**
    actionSet = actionSet ∪ {(connectComponents, S)}
  **return** actionSet
 **else return** *{ Forget }*

**expandState(S, i, action):**
 let $S=((cid_1, deg_1), (cid_2, deg_2), ..., (cid_{nv(i)-1}, deg_{nv(i)-1}))$
 **if** ($D_i$ is an Introduce node) **then**
  **if** (action=newComponent) **then**
   newcid=max{$cid_1, ..., cid_{nv(i)-1}$}+1
   $S'=((cid_1, deg_1), ..., (cid_{nv(i)-1}, deg_{nv(i)-1}), (newcid, 0))$
   **return** (S', $T_{i-1}$[S]+w($v_{i,nv(i)}$), true)
  **else if** (action=(addAsLeaf, $v_{i,j}$)) **then**
   C=w($v_{i,nv(i)}$)
   **if** ($deg_j$=1) **then** C=C-w($v_{i,j}$)
   $S'=((cid_1, deg_1), ..., (cid_j, min\{2,deg_j+1\}), ..., (cid_j,1))$
   **return** (S', $T_{i-1}$[S]+C, true)
  **else if** (action=(connectComponents, SV)) **then**
   **if** ($\exists v_{i,j}, v_{i,k} \in SV . cid_j = cid_k$ ) **then**
    **return** ( (), -Infinity, false)
   newcid= $\max_{v_{i,j} \in SV}$ {$cid_j$}
   $S'=((cid_1,deg_1), ..., (cid_{nv(i)-1}, deg_{nv(i)-1}), (newcid, 2))$
   C=0
   **for** *j=1* **to** *nv(i)-1* **do**
    **if** ($v_{i,j}$ in SV) **then**
     **if** ($deg_j$=1) **then** C=C+w($v_{i,j}$)
     S'[j]=(newcid, min{2, $deg_j$+1})
    **else if** ( $cid_j \in \{cid_k | v_{i,k} \in SV\}$ ) **then**
     S'[j]=(newcid, $deg_j$)
   **return** (S', $T_{i-1}$[S]-C, true)
 **else**
  $v_{i-1,j}$ = the "forgotten" node
  **if** ( $\neg(\exists v_{i-1,k} . k \neq j \land cid_k = cid_j)$ ) **then**
   **return** ((), -Infinity, false)
  $S'=((cid_1,deg_1),...,(cid_{j-1},deg_{j-1}),(cid_{j+1},deg_{j+1}),...,(cid_{nv(i)-1},deg_{nv(i)-1})$
  **return** (S',$T_{i-1}$[S],true)

Using similar state definitions, we can solve other spanning tree problems, like finding a spanning tree with the maximum degree at most Q. In this case, we only need to solve a decision problem (like the first coloring problem we presented) and the degree values in a state S would range from 0 to Q. Of course, we would need to reject states in which the degree of some vertex exceeds Q. Using the decision function in a binary (or linear) search procedure, we can solve the well-known Minimum Degree Spanning Tree problem.

*H. Minimum Weighted Maximal Matching*

A matching is a set of pairs of adjacent vertices, such that each vertex belongs to at most one pair. Given a weight $w(u,v)$ for each edge (u,v) of the graph, the weight of the matching is the sum of the weights of the edges composing it. A matching is maximal if no other edge can be added to the matching. A minimum weighted matching is a maximal matching with minimum weight. In order to find such a matching, we will use states of the form $(s_1, ..., s_{nv(i)})$ for a node $D_i$, where $s_j$ is either 1 or 0 (vertex $v_{i,j}$ belongs to the matching or not). When introducing a vertex at a node $D_i$, the actions are of two types: (*Add, $v_{i,j}$*), meaning that the introduced vertex is added to the matching together with one of its neighbors $v_{i,j}$ with $s_j=0$ and (*Do Not Add*). When forgetting a vertex $v_{i-1,j}$, the algorithm must check that all the neighbors $v_{i-1,k}$ of this vertex in the last node $D_{i-1}$ which contained the vertex are in state $s_j=1$ (to make sure that the matching is maximal). For every valid action (*Add, $v_{i,j}$*), the cost returned by the *expandState* function will be $T_{i-1}[S]+w(v_{i,j}, v_{i,nv(i)})$. For the other actions, the cost does not change ($T_{i-1}[S]$ is returned). We will not present the code, due to lack of space.

*I. Maximum Average Weight Path with Length Constraints*

Given a weight w(u) for each vertex of the graph, we want to find a path whose average weight (sum of the weights of the vertices on the path divided by the number of vertices on the path) is maximum and whose length (number of vertices) is at least L and at most U. The states will have the form $(s_1, ..., s_{nv(i)}, x)$, where:
- $s_j=2$, if $v_{i,j}$ is the only vertex on the path
- $s_j=1$, if $v_{i,j}$ is one of the two end-vertices of the path
- $s_j=0$, if $v_{i,j}$ is not one of the two end-vertices of the path
- x is the total number of vertices on the path (so far)

The set of actions of an Introduce node consists of three actions { Start Path, Select, Do Not Select } and that of a Forget node will be only {Forget}. The framework functions are similar to those from the k-replica placement problem (except that the *better* function selects the maximum value). When selecting a vertex $v_{i,j}$ to be part of the path, we also select an edge from one of the two end-vertices to $v_{i,j}$. As a result, the weight of the path increases by $w(v_{i,j})$, the total number of vertices on the path increases by 1 and the state of the former end-vertex which is now adjacent to $v_{i,j}$ is set to 0. When x=0, we also have the possibility of starting a new path, on which $v_{i,j}$ is the only vertex (so far). Every intermediate state is also a valid state. The state of valid final states is composed of those states S where the number of vertices x has the property $L \leq x \leq U$. From the set of valid final states, we will choose that state $S=(s_1, ..., s_{nv(P)}, x)$ for which $T_P[S]/x$ is maximum. By simply changing the *better* function, we can also solve the minimum weight average path with length constraints problem.

*J. Optimizations for Partial Grid Graphs*

A (m,n) grid graph has m×n vertices arranged on m rows and n columns. Each vertex is adjacent to at most four other vertices (on the rows above and below and the columns to the left and to the right). Such graphs appear, for instance, in processor interconnection networks. A partial (m,n) grid graph is a (m,n) grid graph in which some of the vertices and some of the edges may be missing. These graphs have their pathwidth bounded by min{m,n}. Let's assume that one of the dimensions is bounded by a constant (without loss of generality, we will assume this dimension is n). A path decomposition with pathwidth n can be easily obtained by ordering the vertices from the first to the last row and, for each row, from the first to the last column and introducing (and forgetting) the vertices in this order. Moreover, the vertices within a node $D_i$ are ordered using the same criterion. Partial grid graphs are planar graphs and because of this, the number of states in the dynamic programming algorithm can be reduced by making use of the Catalan property (mentioned in [9]). Let's consider the minimum path (cycle) cover problem. If a state S of a node $D_i$ contains both endpoints of two different paths, then let's assume that $v_{i,a}$ and $v_{i,b}$ (a<b) are the endpoints of the first path and $v_{i,c}$ and $v_{i,d}$ (c<d) are the endpoints of the second path. One of the following conditions must hold: (b<c), (d<a), ((a<c) and (d<b)), ((c<a) and (b<d)). States S which do not obey any of these conditions can be removed.

*K. Covering a Partial Grid Graph with Rectangular Pieces*

We are given a m×n partial grid graph (with n bounded by a constant C) and a set of K types of rectangular pieces { $(r_1,c_1)$, $(r_2,c_2)$, ..., $(r_K,c_K)$ }. The $i^{th}$ type covers $r_i$ rows and $c_i$ columns of the graph (i.e. covers $r_i \times c_i$ vertices). $c_i$ is bounded by n and $r_i$ is bounded by a constant R. We want to place on the graph as many rectangular pieces as possible, under the following restrictions:
- no two rectangular pieces should overlap
- a rectangular piece cannot cover missing vertices
- a rectangular piece must be fully included in the graph
- we can use as many pieces of any type as we want
- we cannot rotate the pieces (although it's not excluded that both $(r_i,c_i)$ and $(c_i,r_i)$ belong to the set)

This problem is more difficult than the previous ones, because it is formulated taking into consideration the information that the graph is a partial grid graph (and not just a graph with bounded pathwidth). The dynamic programming states will have the following form: $(h_1, h_2, ..., h_{nv(i)})$, where $h_j$ is the number of vertices above (and including) the vertex $v_{i,j}$ (i.e. on the same column) which are not covered and not missing from the graph. We are only interested in values $0 \leq h_j \leq R$ (if there are more than R vertices not covered and not missing above $v_{i,j}$, we will limit this value to R). Then, when a new vertex is introduced, the set of actions consists of either not doing anything or choosing to place one of the K types of rectangular pieces with its lower right corner at vertex $v_{i,j}$ (if possible – i.e. for a piece $(r_p, c_p)$ and the vertex $v_{i,j}$ located on row a and column b, we have: $b \geq c_p$, none of the vertices b, b-1, ..., b-$c_p$+1 are missing from row a and they have at least $r_p$ non-covered and non-missing vertices above them). There is an extra problem here, given by the fact that we need to know the value of $h_j$ (for the vertex $v_{i,j}$) at the moment the vertex is introduced. For this, we will consider the pathwidth to be n+1. This way, when $v_{i,j}$ is introduced, the vertex right above $v_{i,j}$ is located in $D_i$ and we can compute $h_j$ as 1 plus the corresponding value for the vertex above $v_{i,j}$ (or just 1 if that vertex is missing from the grid graph). The framework functions are similar to those presented for the other problems.

*L. Other Combinatorial Optimization Problems*

There are many other problems which can be solved using the generic dynamic programming algorithm presented in Section III. Minimum (Weighted) Vertex Cover, Maximum (Weighted) Independent Set, Maximum (Weighted) Matching, Minimum Dominating Set are just a few examples. Some of them can be solved using state definitions and sets of actions very similar to the ones we presented in this section, but for others, the sets of states and actions are more complicated.

## V. RELATED WORK

Graphs with bounded pathwidth and more generally, with bounded treewidth, have been studied intensively during the last two decades, because of their applications in many fields [1, 4, 5, 8]. Frameworks for algorithms on graphs with bounded pathwidth (treewidth) were proposed in [6, 7]. In [6], a framework for solving any problem expressible in monadic second order logic is presented. This framework generates the entire algorithm for solving a problem, but the constant factors of these algorithms are very large. In [7], a framework for network reliability problems based on a model with vertex and edge failure probabilities is presented. Problems related to minimum path and cycle covers were treated in [2,3].

## VI. CONCLUSIONS AND FUTURE WORK

In this paper we presented a dynamic programming framework for solving combinatorial optimization problems on graphs with bounded pathwidth. The framework is general enough to fit many important network optimization problems, which have applications in network reliability analysis and in the improvement of performance and fault tolerance. Compared to other existing frameworks, some problem-specific details, like state definitions and action sets are left to the programmer (solution developer). This makes the development of solutions to problems more difficult, but it also allows for optimizations which might not have been possible otherwise. As future work, we established two directions. The first is related to embedding the solutions of some problems which fit the proposed framework into techniques for improving the performance, fault tolerance and reliability of networks. The second research direction refers to extending the framework to the more general class of graphs with bounded treewidth.